\documentclass[letterpaper, 10 pt, conference]{ieeeconf}  

\IEEEoverridecommandlockouts                              

\usepackage{graphics} 
\usepackage{epsfig} 
\usepackage{mathptmx} 
\usepackage{times} 
\usepackage{amsmath} 
\usepackage{amssymb}  
\usepackage{xcolor}       
\usepackage{epstopdf}
\usepackage{subfigure}
\usepackage{bm}

\title{\LARGE \bf
Equation-Free Coarse Control of Distributed Parameter Systems via Local Neural Operators
}

\author{Gianluca Fabiani$^{1}$, Constantinos Siettos$^{2}$,  and Ioannis G. Kevrekidis$^{3}$
\thanks{*G.F. and I.G.K. work was partially supported by the Department of Energy under Grant No. DE-SC0024162,
as well as partial support by the National Science Foundation under Grants No. CPS2223987 and FDT2436738. C.S. acknowledges partial support by the PNRR MUR projects PE0000013-Future Artificial Intelligence Research-FAIR \& CN0000013 CN HPC - National Centre for HPC, Big Data and Quantum Computing, Gruppo Nazionale Calcolo Scientifico-Istituto Nazionale di Alta Matematica (GNCS-INdAM)}
\thanks{$^{1}$G. Fabiani is with Hopkins Extreme Materials Institute and Department of Chemical and Biomolecular Engineering, Johns Hopkins University, 3400 North Charles Street, Baltimore, 21218 MD, USA
        {\tt\small gfabian2@jh.edu}}%
\thanks{$^{2}$C. Siettos with the Dipartimento di Matematica e Applicazioni "Renato Caccioppoli", Università degli studi di Napoli Federico II, Via Vicinale Cupa Cintia, 26, 80126 Naples NA, Italy
        {\tt\small constantinos.siettos@unina.it}}%
\thanks{$^{3}$I.G. Kevrekidis with Department of Chemical and Biomolecular Engineering and Department of Applied Mathematics and Statistics, Johns Hopkins University, 3400 North Charles Street, Baltimore, 21218 MD, USA
        {\tt\small yannisk@jhu.edu}}%
}

\begin{document}

\maketitle
\thispagestyle{empty}
\pagestyle{empty}

\begin{abstract}
The control of high-dimensional distributed parameter systems (DPS) remains a challenge when explicit coarse-grained equations are unavailable.
Classical equation-free (EF) approaches rely on fine-scale simulators treated as black-box timesteppers. However, repeated simulations for steady-state computation, linearization, and control design are often computationally prohibitive, or the microscopic timestepper may not even be available, leaving us with data as the only resource.
We propose a data-driven alternative that uses \emph{local neural operators}, trained on spatiotemporal microscopic/mesoscopic data, to obtain efficient short-time solution operators.
These surrogates are employed within Krylov subspace methods to compute coarse stable and unstable steady states, while also providing Jacobian information in a matrix-free manner. Krylov-Arnoldi iterations then approximate the dominant eigenspectrum, yielding reduced models that capture the open-loop slow dynamics without explicit Jacobian assembly.
Both discrete-time Linear Quadratic Regulator (dLQR) and pole-placement (PP) controllers are based on this reduced system and lifted back to the full nonlinear dynamics, thereby closing the feedback loop.
\color{black}
The framework is validated by stabilizing an unstable steady-state of the Liouville-Bratu PDE, demonstrating consistent performance between the learned surrogate and the true system, with quantified degradation under plant-model mismatch.
\color{black}
\end{abstract}

\section{INTRODUCTION}
Modelling and, importantly controlling, high-dimensional, multiscale complex systems remains a major challenge across engineering and science~\cite{kevrekidis2003equation, karniadakis2021physics, fabiani2024task}.
Many such systems are governed by complex interactions among large numbers of agents or particles, often leading to emergent behavior that is difficult to predict and regulate, including catastrophic shifts or major irreversible changes in dominant mesoscopic or macroscopic spatio-temporal patterns~\cite{scheffer2003catastrophic}. 
These systems frequently exhibit multistability, with multiple possible steady-states -- some stable, others unstable -- among which certain states may be desirable to reach or maintain through control interventions~\cite{armaou2004time, patsatzis2023data, alvarez2023discrete}.  
%

\begin{figure*}
    \centering
    \includegraphics[trim={2cm 0 1.5cm 0},clip,width=0.95\linewidth]{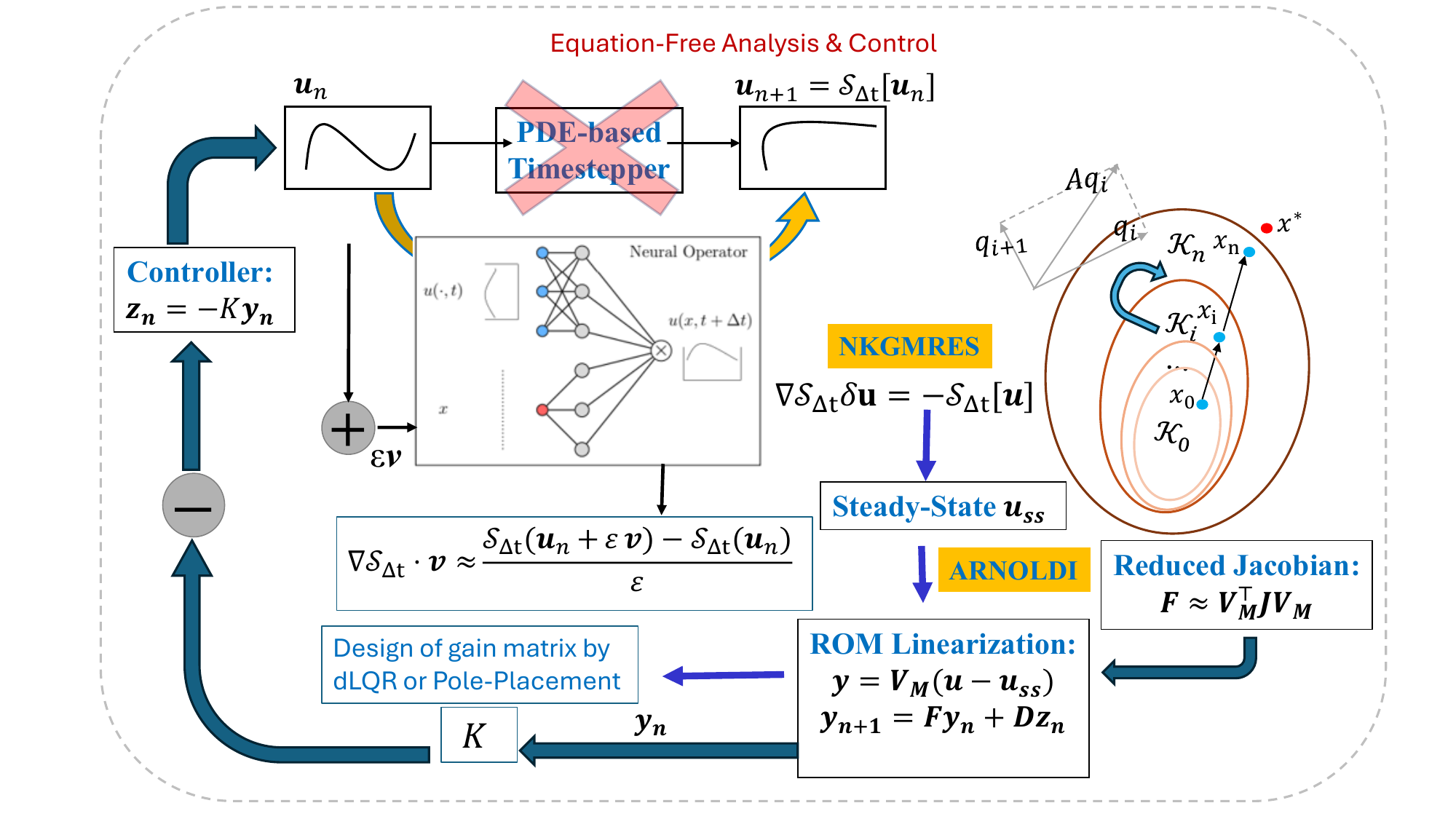}
    \caption{Schematic of the Equation-Free control pipeline via Neural Operators (NOs). An NO, trained on spatiotemporal data, replaces the unavailable traditional PDE-based time-stepper. This surrogate enables matrix-free computation of coarse steady-states (via Newton-Krylov GMRES) and the dominant eigenmodes of the linearization (via Arnoldi iteration). The resulting low-dimensional model is used to design a controller (here, dLQR or pole-placement), which is then implemented in closed loop by feeding control inputs back to the NO-timestepper. \label{fig:schematic}} 
\end{figure*}

The control of high-dimensional distributed parameter systems (DPS) remains especially challenging when explicit coarse-grained equations are unavailable. In such cases, classical control design tools cannot be directly applied, as they typically rely on low-dimensional state-space or PDE-based representations~\cite{christofides2002nonlinear, shvartsman1998low}. 
This motivates frameworks that bypass macroscopic models for enabling principled, data-driven control.
One of such frameworks is the \emph{Equation-Free} (EF) approach~\cite{kevrekidis2003equation, kevrekidis2004equation, erban2007variable}. 
Assuming access to microscopic simulators (such as Agent-Based models or Molecular Dynamics), and under suitable process conditions, EF constructs coarse timesteppers to evolve macroscopic observables directly. These timesteppers can then be embedded in computational superstructures for tasks such as stability and bifurcation analysis, as well as for control design, relying on well-established numerical methods including 
matrix-free Krylov solvers (e.g., Generalized Minimal Residual (GMRES)), and eigensolvers such as the Arnoldi algorithm~\cite{kelley1995iterative, saad2011numerical}.

However, a limitation of the classical EF framework is that its surrogates are constructed \emph{on demand}, requiring repeated calls to the underlying microscopic simulator, which might be computationally prohibitive.
Besides, in many applications the microscopic simulator may not even be available, leaving observational data as the only viable resource. 

In this context, a longstanding goal is to learn effective models or reconstructing operators that describe the system’s evolution, often under partial-information or sparse-data regimes. 
%
\color{black}
Recent data-driven methods learn operators directly from high-fidelity simulations~\cite{raissi2018deep, lee2020coarse, chen2021solving, lu2021learning, vlachas2022multiscale, galaris2022numerical, fabiani2024task}, relaxing the assumptions required by classical analytical approaches~\cite{helfmann2021interacting, deng2025hilbert}.
\color{black}
However, these methods introduce their own challenges: sparse or partial observations; the ``curse of dimensionality''; strong nonlinearities and stiffness; multiple, separated time scales; and 
uncertainty  --  all of which exacerbate ill-posedness.

On top of these difficulties, algorithmic choice becomes critical. Common tools in machine learning -- standard artificial neural networks (ANNs), Gaussian processes (GPs) -- are designed for finite-dimensional regression, i.e., functions $f:\mathbb{R}^n\to\mathbb{R}^m$. By contrast, model discovery for multiscale systems often requires learning operators $\mathcal{G}:\mathrm{U}\to\mathrm{V}$ that map between infinite-dimensional function spaces, for example, the infinite-dimensional nature of models such as PDEs. Neural Operators (NOs)~\cite{lu2021learning,li2020fourier} were introduced to address this gap, yielding surrogates that can generalize across input functions and possibly different resolutions. 
%
Representative NO architectures include DeepONet~\cite{lu2021learning}, Fourier NOs (FNOs)~\cite{li2020fourier}, Graph NOs (GNOs)~\cite{li2020multipole},  
and more recently Random Projection-based Operator Networks (RandONets)~\cite{fabiani2025randonets}. 
For spatio-temporal data, one can learn either the continuous-time RHS governing operator~\cite{lee2020coarse,fabiani2025randonets}, representing the system’s derivative mapping, 
or the solution operator (timestepper), mapping an initial state to a future state over finite time~\cite{li2020fourier, fabiani2025enabling}. 
Both approaches face challenges for long-time predictions, particularly in systems with chaotic behavior, multiple basins of attraction, or high-frequency modes that are difficult to resolve due to the spectral bias of NOs~\cite{li2022learning, wang2023long}. To mitigate this, local-in-time and sometimes local-in-space operators are preferable, trained on short-time snapshots and augmented to maximize data efficiency~\cite{fabiani2025enabling, wang2023long, li2022learning}.

NOs have so far been used mainly to accelerate brute-force simulations, providing fast surrogates once trained. 
\color{black}
Our objective, however, extends beyond forward simulation: we aim to construct neural timesteppers that serve as surrogate black-box operators within an equation-free multiscale framework~\cite{kevrekidis2003equation,kevrekidis2004equation,erban2007variable}, enabling system-level analysis and control without explicit coarse-grained equations.
%

In recent work~\cite{fabiani2025enabling}, we introduced Local NOs as an equation-free building block, coupled with matrix-free Krylov methods~\cite{kelley1995iterative, saad2011numerical}, for coarse-grained bifurcation analysis. Here, we extend this framework to coarse feedback control of dissipative distributed processes, building on equation-free control strategies~\cite{siettos2003coarse,armaou2004time,siettos2006equation,siettos2012equation,patsatzis2023data}.
%
\color{black}
Specifically, unlike~\cite{fabiani2025enabling}, which focused on operator identification and its analysis, this paper introduces a full feedback control pipeline, including reduced-order controller design and closed-loop stabilization.
\color{black}

Our framework follows a two-step coarse controller design. First, fine-scale simulations are processed through the coarse timestepper to identify the target coarse stationary state, approximate the subspace spanned by the dominant slow modes in its neighborhood, and characterize the action of the coarse linearization on this subspace. For this task, we employ Newton-Krylov iterations in combination with the coarse timestepper to construct coarse fixed-point solvers and eigensolvers. This procedure yields an approximate reduced-order coarse model suitable for feedback design.

To enable feedback stabilization, spatially localized actuators are introduced, and their effect is quantified by perturbing the timestepper and recording directional derivatives. This allows us to construct reduced discrete-time input-output models coupling actuator actions with the unstable modes. On these reduced models, we design discrete-time controllers -- both pole-placement (PP) controllers and optimal strategies based on the discrete-time Linear Quadratic Regulator (dLQR)~\cite{armaou2004time, siettos2006equation,ogata1995discrete}. 
Krylov subspace techniques play a central role in this pipeline, enabling the computation of coarse fixed points and matrix-free Jacobian information around equilibria, even for large-scale systems.
Fig.~\ref{fig:schematic} provides a schematic illustration of the EF Analysis and Control framework via local NOs.
We demonstrate the methodology on one representative problem: the parabolic Liouville-Bratu equation~\cite{boyd1986analytical}. We compare results obtained with a NO-timestepper against those obtained with a finite-difference PDE solver. The data-driven and physics-based approaches produce consistent stabilizing feedback laws.

\color{black}
\section{Preliminaries}

\color{black}
\subsection{Problem Statement}
Our objective is to stabilize or steer the dynamics of a distributed parameter system (DPS) via feedback, using only data-driven approximations of its solution operator (time-stepper). We consider a DPS governed by a continuous-time PDE whose exact coarse-grained model is assumed unavailable.

Let $u(\cdot,t) \in \mathrm{U}$ denote the system state at time $t$, where $\mathrm{U}$ is an appropriate function space. The discrete-time solution operator $\mathcal{S}_{\Delta t}: \mathrm{U} \to \mathrm{U}$ advances the state by a time step $\Delta t$:
\begin{equation}
u(\cdot,t+\Delta t) = \mathcal{S}_{\Delta t}[u(\cdot,t)],
\end{equation}
implicitly incorporating boundary conditions. In the equation-free setting, this operator is not available in closed form; instead, we learn a neural operator (NO) surrogate from short-time trajectory data.

Unlike the classical EF approach, where timesteppers are constructed \emph{on demand} and discarded, our NO-based timestepper is trained once and reused for analysis, model reduction, and, crucially, for the design and implementation of feedback controllers.

The proposed control framework is architecture-agnostic and can be implemented with any neural operator capable of approximating the solution operator $\mathcal{S}_{\Delta t}$.
However, in our experiments we employed RandONets~\cite{fabiani2025randonets}, a computationally efficient variant of DeepONets, to learn data-driven approximations of the solution operator $\mathcal{S}_{\Delta t}$.

RandONets use randomized embeddings in place of fully trainable hidden layers, reducing training to a convex least-squares problem and enabling fast, accurate surrogate construction.
In particular, RandONets~\cite{fabiani2025randonets} attempt to address common issues of Deep Learning approaches, related to slow convergence, overparameterization, and high computational cost~\cite{froese2023training, fabiani2025random, karumuri2024efficient}, by replacing fully trainable hidden layers with fixed randomized embeddings.
One way of accomplishing this is by leveraging random projections and the Johnson-Lindenstrauss lemma~\cite{johnson1984extensions} and exploiting the universal approximation power of random features~\cite{igelnik1995stochastic, rahimi2007random, rahimi2008uniform, fabiani2023parsimonious}.
\emph{This fixed randomized architecture design converts the original non-convex, gradient-based training of DeepONets into a convex linear least-squares one}~\cite{fabiani2025randonets}. \color{black}

\subsection{Equation-Free system-level computations via local NOs: fixed-points and stability}
Learning long-time prediction 
is challenging due to stiffness and spectral bias. It is often more efficient to learn local-in-time solution operators $\mathcal{S}_{\Delta t}$ and apply them autoregressively, taking advantage of the semigroup property, to reconstruct a long time $T$ horizon predictor:
\begin{equation}
u(\bm{x},T) = \mathcal{S}_T[u]=\underbrace{\mathcal{S}_{\Delta t} \circ \mathcal{S}_{\Delta t} \circ \cdots \mathcal{S}_{\Delta t}}_{\frac{T}{\Delta t} \text{times}}[u].
\end{equation}
This can suffer from error accumulation, especially for high-frequency modes. Instead, steady-states can be obtained directly as fixed points of the short-time solution operator:
\begin{equation} \label{eq:NO_timestepper}
u^* = \mathcal{S}_{T}[u^*], \qquad
\psi(u) = u - \mathcal{S}_{T}[u].
\end{equation}
Newton’s method computes zeros of $\psi$:
\begin{equation} \label{eq:newton}
\nabla\psi(u^{(k)})\delta^{(k)} = -\psi(u^{(k)}), \quad
\delta^{(k)} = u^{(k+1)}-u^{(k)}.
\end{equation}

\subsection{Krylov Subspace Methods}
Krylov methods are among the top ten algorithms of the 20th century for a reason: they solve large linear systems $(Ax=b)$ efficiently using only matrix-vector products. Starting from $x^{(0)}$ with initial residual $r^{(0)}=b-Ax^{(0)}$, Krylov methods build iteratively an increasing family of subspaces $\mathcal{K}_i$ and a sequence of approximations $x_i$:
\begin{equation}
x_1, x_2, \dots, x_n \in x_0 + \mathcal{K}_i(A,r_0), \quad i=1,\dots,n,
\end{equation}
where the \emph{Krylov subspace} is defined as
\begin{equation}
\mathcal{K}_n = \text{span}\{ r_0, Ar_0, A^2 r_0, \dots, A^{n-1} r_0 \}.
\end{equation}
In GMRES the approximation $x_i$ is chosen to minimize the residual norm over that affine space:
\begin{equation}
    x_i=arg\hspace{-0.6cm}\min_{x\in x_0+\mathcal{K}_i(A,r_0)}\|b-Ax\|_2.
\end{equation}
Convergence depends on the spectral distribution of $A$; preconditioning and restarting are commonly used in practice.
For large nonlinear systems $\psi(u)=0$, Newton-Krylov methods solve the linearized system $J^{(k)}\delta=-\psi(u^{(k)})$ with $J^{(k)}=\nabla\psi(u^{(k)})$.
The Jacobian-vector products needed in this process can be cheaply obtained via directional finite difference approximation:
\begin{equation}
\nabla\psi(u^{(k)}) r_0 \approx \frac{\psi(u^{(k)}+\epsilon r_0)-\psi(u^{(k)})}{\epsilon}.
\end{equation}
Around the obtained steady-state, we compute approximate critical eigenvalues and eigenvectors via the Krylov-Arnoldi algorithm. Arnoldi builds an orthonormal basis $Q_m$ of the Krylov subspace $\mathcal{K}_m$. The Jacobian projected onto this basis yields an upper Hessenberg matrix $H_m = Q_m^\top J Q_m$. The Ritz values (eigenvalues of $H_m$) approximate the critical eigenvalues of the full Jacobian, thus providing stability information~\cite{saad2011numerical}.
The associated Ritz vectors $Q_m v$ span a low-dimensional subspace $\mathcal{P}$ approximating the dominant eigenspace of the Jacobian, which can be used to analyze and locally linearize the system~\cite{ogata1995discrete}.

\section{Methods}
In this section, we outline the construction of linear discrete-time controllers for spatially distributed processes using NO-timestepper surrogates. The procedure consists of two stages. First, the surrogate is employed-together with Newton-Krylov GMRES and Arnoldi iterations to identify the dominant coarse dynamics, approximate coarse fixed points, and estimate the associated slow Jacobian. Second, based on this reduced linearized model, we design discrete-time controllers using either pole placement or optimal strategies such as dLQR.

\subsection{Controlled Neural Operator Timestepper}
The dynamics of the spatially distributed process are represented by a NO-timestepper $\mathcal{S}_{\Delta t}$, which advances the state $u_n$ by one step of size $\Delta t$.  
We introduce the (open-loop) controlled NO-timestepper
\begin{equation}
u_{n+1} = \Phi(u_n, z_n) = \mathcal{S}_{\Delta t}(u_n) + G(u_n, z_n),
\label{eq:controlled_8}
\end{equation}
where $u_n = u(\cdot,t)$, $u_{n+1} = u(\cdot,t+\Delta t)$, and $z_n \in \mathbb{R}^k$ denotes the control input. The actuator contribution $G(u_n, z_n)$ satisfies $G(u_n,0) \equiv 0$ and represents the correction induced by the applied controls.  
In the ideal setting, $\Phi$ constitutes the fully controlled one-step-ahead operator, combining the autonomous dynamics $\mathcal{S}_{\Delta t}$ with the actuator effects, and requires training with data that include open-loop controlled trajectories. 

In many practical settings (see Section~\ref{sec:Bratu}), only autonomous, uncontrolled spatio-temporal trajectories are available, allowing us to train only the neural timestepper $\mathcal{S}_{\Delta t}$. Training the full open-loop timestepper $\Phi$ directly, or the residual contribution $G$ in Eq.~\eqref{eq:controlled_8}, in a NO framework would require significantly more data than the autonomous case, as each control actuator defines a parametric family of operators. 
Besides, such an approach is tied to a specific actuator design and would require retraining if the design is modified.
Therefore, when controlled training data are unavailable, we must infer the effect of actuation through alternative means. We consider two complementary approaches:
\paragraph{Known actuator physics (model-based)} 
If the actuator physics are known from first principles or design specifications  --  including their spatial distribution and how they couple with the state  --  we can incorporate this knowledge directly.
Let us assume that in continuous time, the controlled dynamics take the general form
\begin{equation}
\frac{\partial u}{\partial t} = \mathcal{F}(u) + \mathcal{C}(u, z),
\end{equation}
where \(\mathcal{F}\) represents the autonomous dynamics and \(\mathcal{C}\) captures the actuation. 
Over a sufficiently small timestep \(\Delta t\), a first-order discrete approximation yields
\begin{equation}
u_{n+1} \simeq \mathcal{S}_{\Delta t}(u_n) + \Delta t \, \mathcal{C}(u_n, z_n).
\label{eq:controlled_nonlinear}
\end{equation}
This approximation retains the leading-order effect and is compatible with our goal of learning short-time operators \(\mathcal{S}_{\Delta t}\).
No controlled training data are required in this scenario, as $\mathcal{C}$ is derived from design knowledge.
This formulation can accommodates various actuation mechanisms, including linear body forcing \(\mathcal{C}(u_n, z_n) = B z_n\) with fixed spatial profiles (as in Section~\ref{sec:Bratu}). 


\paragraph{Data-driven local probing}
When the actuator physics are unknown due to complex actuation mechanisms, we can, if experimentation is feasible, estimate the local linearized actuator response through targeted perturbation experiments around a desired operating point. These experiments/simulations are designed specifically to probe the system's sensitivity to actuation in the neighborhood of the target state, yielding a local linear approximation. 
Specifically, we select a reference state $u^*$ 
and apply small impulsive inputs $z = \epsilon \bm{e}_j$ for each actuator $j = 1,\dots,k$. Observing the true system response over one timestep yields a finite-difference approximation of the actuator Jacobian:
\begin{equation}
H_{:,j} \approx \frac{\mathcal{T}(u^*, \epsilon \bm{e}_j) - \mathcal{T}(u^*, 0)}{\epsilon},
\label{eq:probing}
\end{equation}
where $\mathcal{T}$ denotes the true (unknown) open-loop control dynamics, accessed experimentally or via a high-fidelity physics-based simulator.
This matrix $H \in \mathbb{R}^{N \times k}$ captures the local linearized effect of actuation. 
This approach requires only $k+1$ short trajectories (one per actuator plus the uncontrolled reference), making it feasible even in data-limited settings. Moreover, it naturally accommodates complex actuation mechanisms   --   including boundary control or nonlinear input effects   --   as it directly measures the system's response rather than relying on a prescribed functional form.

Both approaches enable the construction of an open-loop surrogate 
without requiring full controlled training data. 

\subsection{Obtaining the Coarse Slow Linearization}
The system identification stage aims to determine the target coarse stationary state $u_{ss}$ and a linearized model of the slow dynamics around it. This is accomplished by first identifying the autonomous slow behavior and then evaluating the effect of actuators on the system.

Open-loop stationary states are located as the solution to $u_{ss} = \mathcal{S}_{\Delta t}(u_{ss})$ using a Newton-Krylov-GMRES iteration wrapped around the NO timestepper. In this framework, the Jacobian-vector products are computed in a matrix-free manner, without explicitly Jacobian construction.
The action of the Jacobian of the timestepper at the fixed point, $J = \partial \mathcal{S}_{\Delta t} / \partial x \big|_{(u_{ss})}$, in selected directions is approximated through directional finite differences approximation of matrix-vector product:
\begin{equation}
Jv \approx \frac{\mathcal{S}_{\Delta t}(u_{ss} + \epsilon v) - \mathcal{S}_{\Delta t}(u_{ss})}{\epsilon},
\end{equation}
where $\epsilon$ is a small perturbation parameter \textcolor{black}{(e.g., scaled relative to machine precision $\epsilon_{\text{mach}}$ to balance truncation and round-off error. Here, $\epsilon = \epsilon_{\text{mach}}^{1/3} \|u_{ss}\|$)} and $v$ is an arbitrary vector.
The Arnoldi algorithm provides an orthonormal basis $V_M \in \mathbb{R}^{N \times M}$ for the slow subspace $\mathcal{P}$, and a low-dimensional representation of the Jacobian's action in this subspace, $F = V_M^\top J V_M \in \mathbb{R}^{M \times M}$.

This yields a low-dimensional linearization of the slow subsystem. Let $y_n \in \mathbb{R}^M$ be the coordinates of the deviation from the stationary state within the slow subspace, defined by the projection:
\begin{equation}
y_n = V_M^\top (u_n - u_{ss}).
\end{equation}
The dynamics of these slow coordinates are approximated by the uncontrolled discrete-time system:
\begin{equation}
y_{n+1} = F y_n, \quad y_n \in \mathbb{R}^M,
\end{equation}

{\bf Actuator Effects.} Once the stationary state is identified, the effect of actuators {\em in the slow subspace} is estimated by perturbing the system and evaluating derivatives with respect to the control inputs. Combining the autonomous linearization with these derivatives yields the low-dimensional approximate open-loop discrete-time system:
\begin{equation}
y_{n+1} = F y_n + D z_n,
\label{eq:reduced_open_loop}
\end{equation}
where $z_n \in \mathbb{R}^k$ represents the control inputs, and $D \in \mathbb{R}^{M \times k}$ approximates the actuator influence on the slow modes.

The matrix $D$ is obtained by projecting the actuator sensitivity $H \in \mathbb{R}^{N \times k}$ onto the dominant slow subspace:
\begin{equation}
D = V_M^\top H,
\label{eq:D_computation}
\end{equation}
where $V_M \in \mathbb{R}^{N \times M}$ spans the slow subspace. The sensitivity matrix $H \in \mathbb{R}^{N \times k}$ captures the local linearized effect of actuation on the full-state dynamics and can be obtained in different ways, depending on the available information:

\begin{itemize}
    \item If a full controlled timestepper $\Phi(u_n, z_n)$ is available, we compute $H$ via finite differences, as in Eq.\eqref{eq:probing}, with $\Phi$ substituting $\mathcal{T}$.
    
    \item If the actuation mechanism is known from first principles, $H$ follows directly from the discrete approximation in Eq.~\eqref{eq:controlled_nonlinear} as $H = \Delta t \, \partial \mathcal{C}/{\partial z}\big|_{(u_{ss},0)}$. 
    
    \item If the actuator physics are unknown but experimentation is feasible, we estimate $H$ by applying small impulsive inputs to the true system $\mathcal{T}$ around $u_{ss}$ (see Eq.~\eqref{eq:probing}).
\end{itemize}
In our experiments, we use  the model-based approach: for our Liouville-Bratu illustrative problem in Section~\ref{sec:Bratu} the actuation mechanism is given in the model.

\subsection{Feedback control design}
Linear feedback controllers are designed to stabilize (or improve the stability characteristics) of the target stationary state exploiting the low-dimensional model. The goal is to design a feedback gain matrix $K \in \mathbb{R}^{k \times M}$ such that the control law $z_n = -K y_n$
stabilizes the target stationary state by shaping the dynamics of the reduced slow subsystem:
\begin{equation}
y_{n+1} = (F - D K) y_n.
\end{equation}
Here, we employ two alternative methodologies to compute a gain matrix $K$.

\paragraph{Pole Placement}
This approach directly assigns the eigenvalues $\mu_{i}$ of the closed-loop matrix $(F - D K)$ to desired locations inside the unit circle. The gain matrix $K$ is the solution to the eigenvalue assignment problem. 
This provides direct control over the transient response of the slow modes.
In practice, we compute $K$ using MATLAB’s \texttt{place} function from the Control Toolbox.

\color{black}
\begin{figure*}[th!]
\centering
    \begin{minipage}{0.92\linewidth}
    \centering
    \subfigure[unstable steady-state]{
    \includegraphics[width=0.31\linewidth]{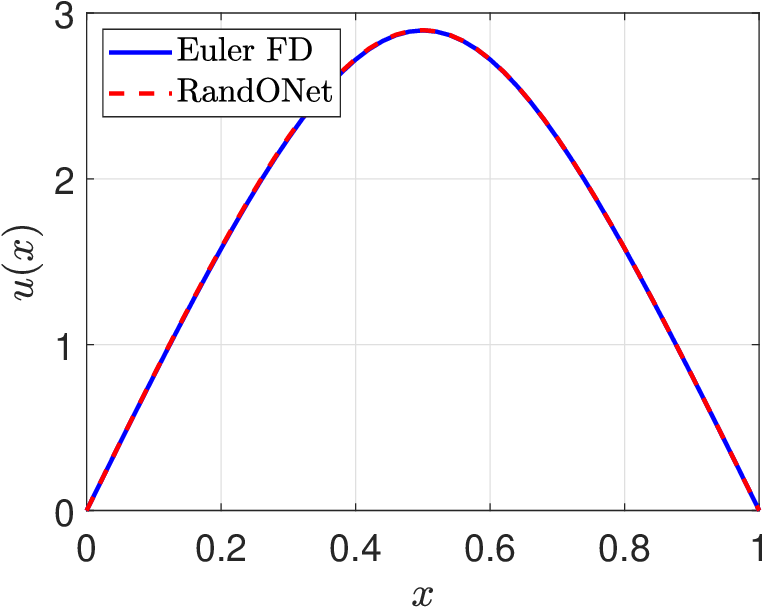}
    }
    \hfill
    \subfigure[open-loop eigenvalues]{
    \includegraphics[width=0.31\linewidth]{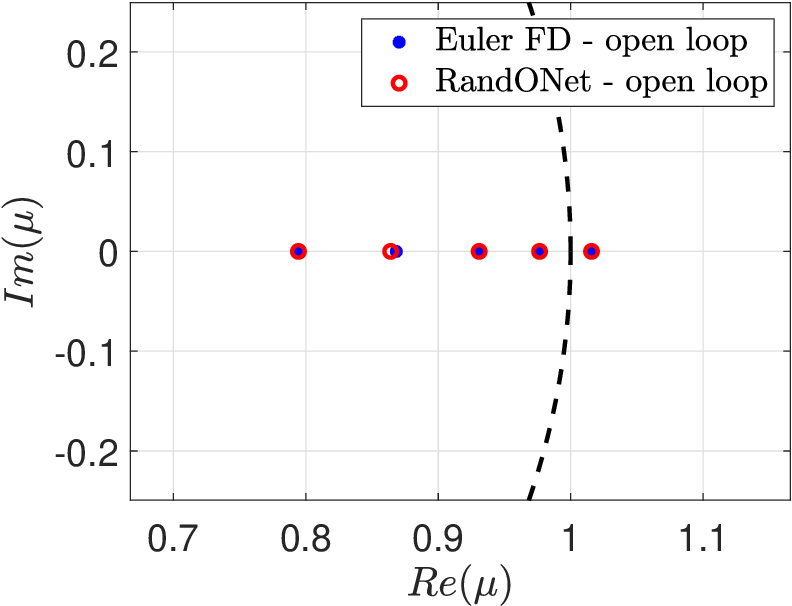}
    }
    \hfill
    \subfigure[close-loop eigenvalues]{
    \includegraphics[width=0.31\linewidth]{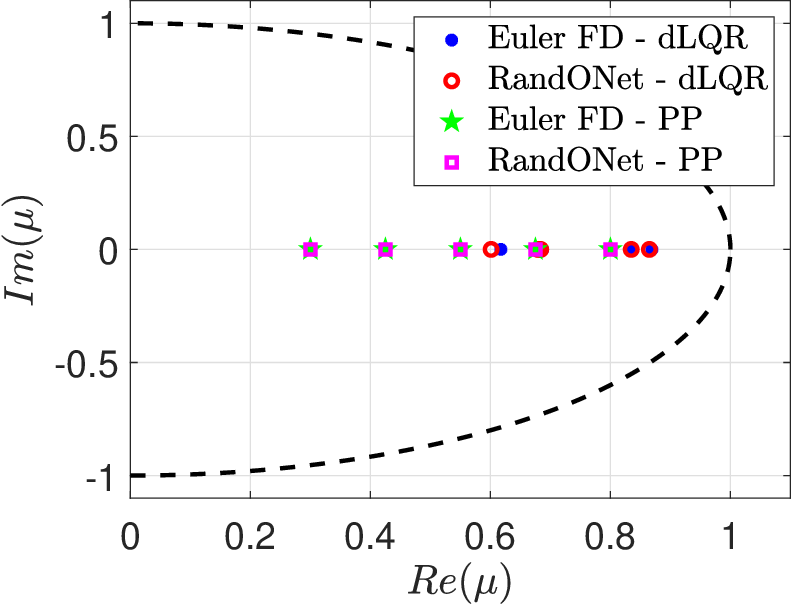}
    }\\
    \subfigure[dLQR - $L^2$-convergence to steady-state]{
    \includegraphics[trim={0 0 0cm 0.2cm},clip,width=0.31\linewidth]{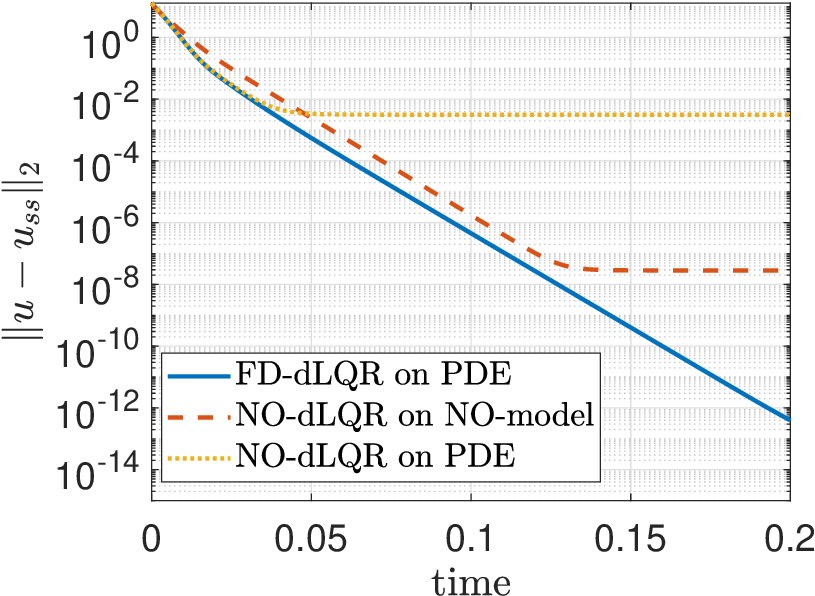}
    }
    \hfill
    \subfigure[NO-dLQR on NO-model: convergence]{
    \includegraphics[width=0.31\linewidth]{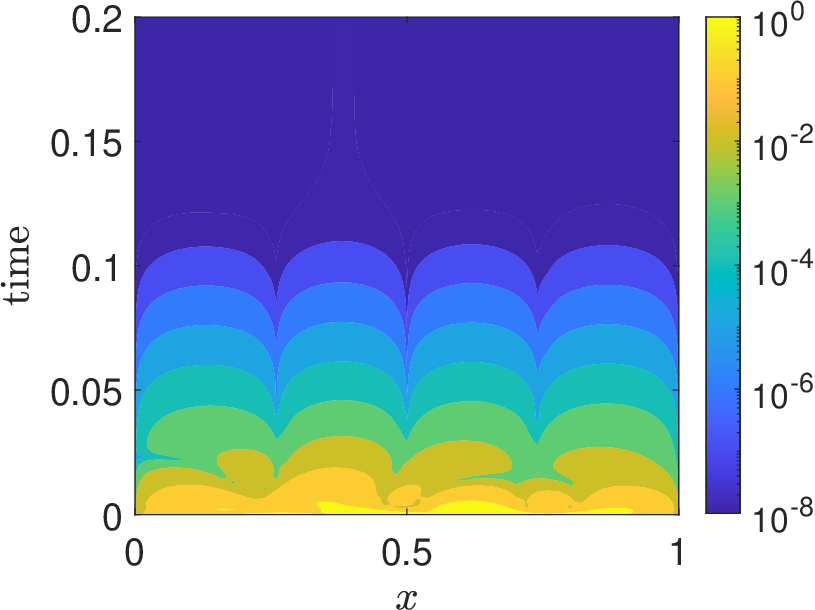}
    }
    \hfill
    \subfigure[NO-dLQR on NO-model: actuator inputs]{
    \includegraphics[trim={0 0 1.1cm 0.5cm},clip,width=0.31\linewidth]{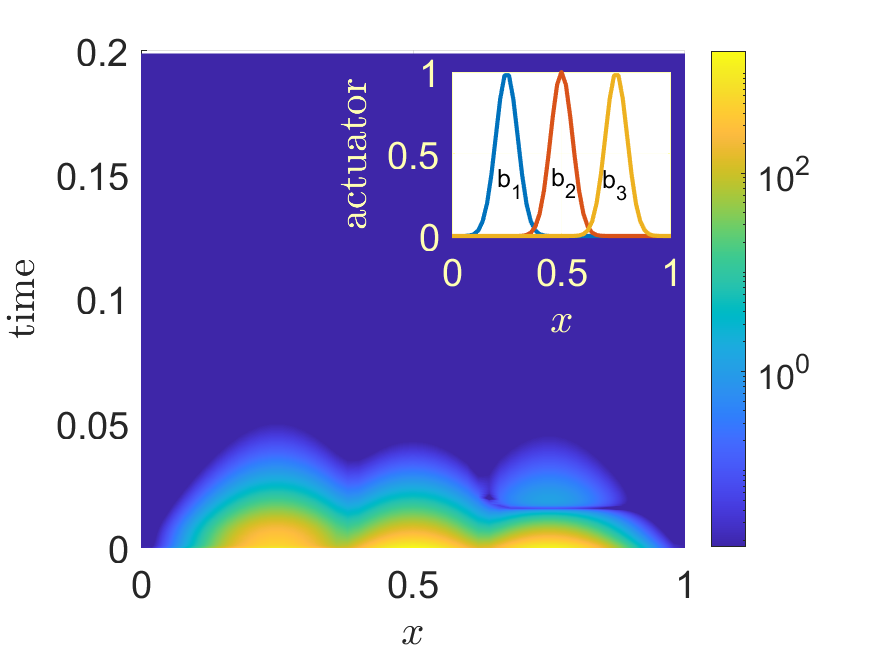}
    }
    \\
    \subfigure[PP - $L^2$-convergence to steady-state]{
    \includegraphics[trim={0 0 0cm 0.2cm},clip,width=0.31\linewidth]{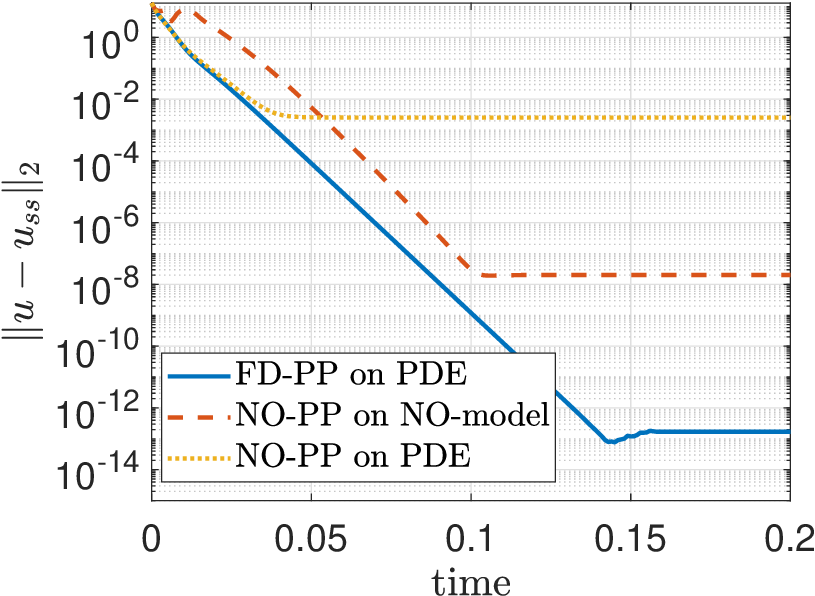}
    }
    \hfill
    \subfigure[NO-dLQR on PDE: convergence]{
    \includegraphics[width=0.31\linewidth]{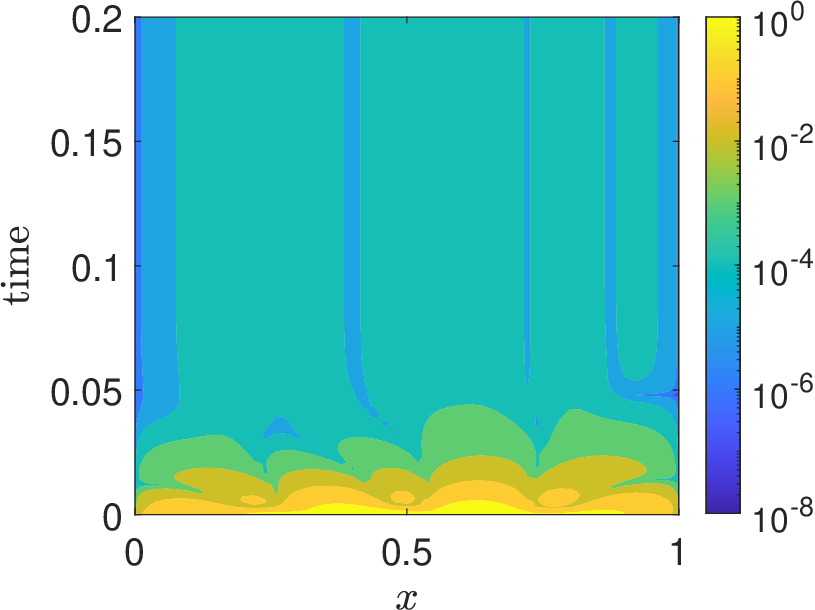}
    }
    \hfill
    \subfigure[NO-dLQR on PDE: actuator inputs]{
    \includegraphics[trim={0 0 1.1cm 0.5cm},clip,width=0.31\linewidth]{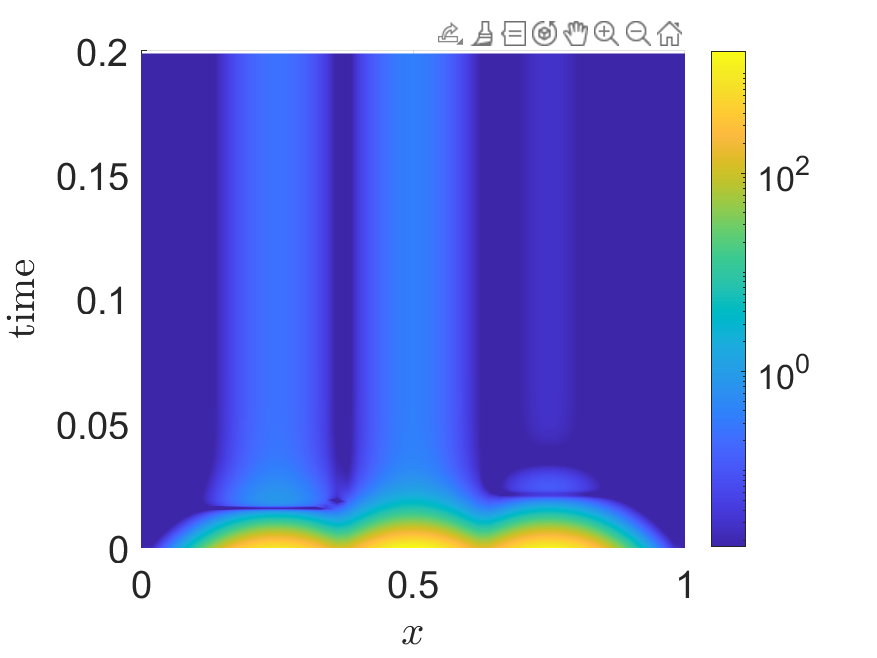}
    }
    \end{minipage}
  \caption{ \color{black}
  Equation-free coarse stabilization of an unstable steady-state of the Liouville-Bratu PDE, in Eq.\eqref{eq:Bratu_parabolic_controlled}, with $\lambda=2$. We use a RandONet-based NO-timestepper and compare it with an Euler-Finite Difference one, based on the actual PDE. (a) The target unstable steady-state $u_{ss}(x)$. (b) Leading eigenvalues $\mu_i$ of the open-loop system's coarse linearization, computed via Arnoldi iterations. (c) Closed-loop eigenvalues after applying the Pole-Placement (PP) and the dLQR controllers. (d) and (g) Convergence of the $L^2$-norm error $||u - u_{ss}||_2$ for three control scenarios: FD on PDE (reference), NO on NO (ideal surrogate), and NO on PDE (model mismatch). Results are shown for dLQR (d) and pole placement (g). (e, h) State evolution $u(x,t)$ using the RandONet-based dLQR controller, shown for the NO surrogate (e) and the original PDE plant (h), illustrating stabilization under model mismatch. (f) and (i) Corresponding absolute amplitude of the spatio-temporal control inputs $|Bz_n|$ of the dLQR controller; in the inset of (f) the spatial profiles of the 3 Gaussian actuators $b_1,b_2$ and $b_3$. 
  \color{black}
  \label{fig:Bratu_results}}
\end{figure*}

\color{black}
\paragraph{Discrete-Time Linear Quadratic Regulator (dLQR)}
The dLQR approach computes the optimal gain matrix $K$ by minimizing the quadratic cost function:
\begin{equation}
\mathcal{J} = \sum_{n=0}^{\infty} \left( y_n^\top Q y_n + z_n^\top R z_n \right),
\end{equation}
where $Q \succeq 0$ and $R \succ 0$ are state and control weighting matrices, respectively. The optimal gain $K$ is obtained by solving the discrete-time algebraic Riccati equation\cite{armaou2004time}. In practice, we compute $K$ using MATLAB’s \texttt{dlqr} function from the Control Toolbox.
\begin{equation}
P = F^\top P F - F^\top P D (R + D^\top P D)^{-1} D^\top P F + Q.
\end{equation}

\section{Numerical Results}
We present numerical results for a benchmark problem: the parabolic Liouville-Bratu equation, using the same learned surrogate timestepper NO model introduced in~\cite{fabiani2025enabling}. 
\color{black}
To assess performance under realistic plant-model mismatch, we test the NO-based controller in two scenarios:
(i) applied to the NO-plant itself---an ideal surrogate scenario---and 
(ii) applied to the high-fidelity PDE solver (finite-difference with forward Euler, $\Delta t_{FD} = 5\times10^{-5}$), which serves as the ground-truth system.
The controller sampling period is $\Delta t = 0.001$ in both cases, matching the timestep of the learned NO.
The performance degradation observed in case (ii) stems from two main sources of uncertainty:
(a) approximation errors in the learned solution operator (trained only on uncontrolled trajectories), and 
(b) incomplete knowledge of the actuator's effect on the true dynamics, since the NO was not trained with controlled data.
This comparison quantifies the practical limitations of data-driven surrogates when deployed in closed loop, while the NO-plant case illustrates near-ideal performance under accurate modeling.
Extensions that incorporate controlled training data, uncertainty quantification, or robust control design are left for future work.
\color{black}

\subsection{The parabolic Liouville-Bratu PDE}
\label{sec:Bratu}
We consider the parabolic Liouville-Bratu PDE~\cite{boyd1986analytical, fabiani2021numerical}, which models a variety of physical and chemical processes:
\begin{equation}
    \frac{\partial u}{\partial t}=  \frac{ \partial^2 u}{\partial x^2}+\lambda \exp(u)+\bm{b}(x)\cdot \bm{z}(t), \qquad x\in \Omega=[0,1],
    \label{eq:Bratu_parabolic_controlled}
\end{equation}
where we fix the parameter $\lambda=2.0$, $\bm{z}(t)\in\mathbb{R}^3$ represent the actuator amplitudes and homogeneous Dirichlet boundary conditions $u(0,t) = u(1,t) =0$ are imposed.
In the PDE in Eq.~\eqref{eq:Bratu_parabolic_controlled}, for actuation, we have considered three spatially localized Gaussian bumps $b_i(x) = \exp(-(x - c_i)^2 / (2\sigma^2))$ with centers $c_i = \{0.25, 0.5, 0.75\}$ and standard deviation $\sigma=0.05$.

The corresponding autonomous one-dimensional steady-state problem admits an analytical solution~\cite{fabiani2021numerical}:
\begin{equation}
\begin{split}
    u(x)=2\ln\frac{\cosh{\theta}}{\cosh{\theta (1-2x)}}; \qquad
    \text{with } \cosh{\theta} = \frac{4\theta}{\sqrt{2 \lambda}}.
\end{split}
\label{eq:sys1}
\end{equation}
For $0<\lambda <\lambda_c$, there exist two solution branches that merge at the saddle-node $\lambda_c \sim 3.513830719$, where stability changes; beyond $\lambda_c$ no steady solutions exist~\cite{boyd1986analytical, fabiani2021numerical}.

In~\cite{fabiani2025enabling}, we trained the RandONet timestepper models, on autonomous synthetic generated data (PDE solutions), using a $m=51$-point spatial discretization for the function input in the branch, and compared the predictions against solutions from a simpler finite-difference/forward Euler scheme, with the same discretization and a 10-times smaller time-step $\Delta t_{FD}=0.0001$.
Full details of the discretization, solver, and dataset structure are provided in~\cite{fabiani2025enabling}, where the complete learned NO and the reconstructed bifurcation diagram are reported.

\textbf{The controlled PDE model.}
Here, we build on our construction of the learned NO to now perform control tasks.
The controller sampling time and the reporting horizon of the neural timestepper are both set to $\Delta t=0.001$.

In practice, we do not have access to a fully trained controlled (``closed loop") timestepper $\Phi(u_n,z_n)$ -- that form of NO construction is the subject of current research.
Instead, we adopt the model-based approximation
\begin{equation}
    u_{n+1} \approx \mathcal{S}_{\Delta t}(u_n) + \Delta_t B z_n,
    \label{eq:controlled_discrete}
\end{equation}
where $B\in\mathbb{R}^{m\times 3}$ corresponds to the spatial actuator configuration in Eq.~\eqref{eq:Bratu_parabolic_controlled}, which is also illustrated in the inset of Figure~\ref{fig:Bratu_results}(f).
To validate our approach, we compare the NO-based controller against a high-fidelity timestepper where the controller is applied directly to the actual PDE. This timestepper employs a finite-difference (FD) scheme with forward Euler integration, using an inner time-step of $\Delta t_{FD}=5\times10^{-5}$ within the outer control step of $\Delta t=0.001$. This provides access to the true controlled timestepper $u_{n+1} = \Phi_{FD}(u_n,z_n)$, where the actuator effects are properly integrated into the dynamics.

For both timesteppers, we design two controllers based on the reduced-order open-loop model $(F,D)$ described in Eq.~\eqref{eq:reduced_open_loop}: a discrete-time Linear Quadratic Regulator (dLQR) with weight matrices $Q=0.5 I$ and $R=dt^210 I$, and a pole-placement controller with desired eigenvalues $\mu_i = [0.30, 0.425, 0.55, 0.675, 0.80]$.

The numerical results are reported in Fig.\ref{fig:Bratu_results}.
We compute the unstable steady-state $u_{ss}(x)$ using a Newton-Krylov-GMRES method, as shown in Fig.~\ref{fig:Bratu_results}(a), for each timestepper. The leading open-loop eigenvalues of the linearized system, computed around $u_{ss}$ via a matrix-free Arnoldi algorithm, are shown in Fig.~\ref{fig:Bratu_results}(b). 
The resulting closed-loop eigenvalues for both the FD-Euler and RandONet models are shown in Fig.~\ref{fig:Bratu_results}(c), quantitatively confirming successful stabilization.

Starting from a perturbed initial condition $u_0 = u_{ss} \cdot \left(1.2 + 0.4 \sin(10\pi x) + 0.4 e^x\right)$, we simulate the closed-loop response. Figs.~\ref{fig:Bratu_results}(d) and (g) show the $L^2$-norm error $\|u - u_{ss}\|_2$ over time. 
\color{black}
Panel (d) corresponds to the dLQR controller, and panel (g) to the pole-placement (PP) controller.
For each controller, three scenarios are compared: 
the NO-based controller applied to the NO-plant (ideal surrogate, dashed), 
applied to the high-fidelity PDE-plant (practical case with model mismatch, dash-dot), 
and---for reference---the same controller applied in a classical FD-based simulation on the PDE (solid). Indicatively, the spatiotemporal evolution of the state under the RandONet-dLQR controller is shown in Fig.~\ref{fig:Bratu_results}(e) for the NO-plant and in Fig.~\ref{fig:Bratu_results}(h) for the PDE-plant, illustrating smooth convergence in the ideal case and the effect of plant-model mismatch in the practical deployment.
The latter case shows mild performance degradation due to uncertainties in the learned dynamics and actuator effect, yet the controller still achieves stabilization. The corresponding absolute amplitude control signals $|B(x)\cdot z(t)|$ are shown in panels (f) and (i) of Fig.~\ref{fig:Bratu_results}(f).

The results show consistent stabilization performance across all tested scenarios. 
The FD-based controller achieves near-machine-precision error when applied to the PDE plant (reference case), while the same NO-based controller applied to the NO-plant saturates at an error of $\sim 10^{-7}$, reflecting the accuracy limits of the learned surrogate.
When deployed on the PDE plant, the NO-based controller exhibits a residual error of $\sim 10^{-3}$, highlighting the effect of plant-model mismatch arising from uncertainties in the learned dynamics and actuator approximation.
This offset arises primarily from the $\sim 2.6\%$ steady-state discrepancy between the learned NO and the true PDE.
Consequently, the control input does not vanish at steady state (Fig.~\ref{fig:Bratu_results}(i)), maintaining a small but persistent actuation to counteract the drift induced by tracking an inaccurate setpoint.
Given the well-conditioned actuator matrix ($\sigma_{\min}(B) \approx 2.10$), the linear scaling estimate $\|u_{ss}^{\text{FD}} - u_{ss}^{\text{NO}}\|_2 / \sigma_{\min}(B) \approx 0.0124$ provides an approximate upper bound on the steady-state offset. The observed residual ($\sim 2.5 \times 10^{-3}$) lies below this bound, confirming that the closed-loop performance is consistent with the level of plant--model mismatch.
%
Future work will focus on improving robustness through controlled-data training, uncertainty quantification, and adaptive compensation strategies.
\color{black}

\section{Conclusion}
In this work, we have revisited the equation-free (EF) framework for the systematic analysis and control of complex systems, addressing a central limitation of the original methodology: its reliance on repeated, potentially expensive, calls to a fine-scale simulator. We have shown that {\em local neural operator (NOs) surrogates}, trained on spatiotemporal data, can effectively replace the need of a first-principles-derived microscopic timestepper. This data-driven NO-timestepper seamlessly integrates into established equation-free computational protocols, enabling the matrix-free identification of coarse steady-states, the estimation of their stability through Arnoldi-based eigensolvers, and the subsequent design of low-dimensional, stabilizing controllers. The proposed approach was demonstrated by successfully stabilizing an unstable, spatially structureed steady-state of the Liouville-Bratu PDE using both dLQR and pole-placement (PP) techniques designed on a reduced-order model, with performance consistent with a controller derived from the known PDE.

This work illustrates an end-to-end equation-free control pipeline requiring only short-horizon simulation data.
\color{black}
Specifically, performance matched that of a controller designed from the known PDE when the same surrogate was used as the plant, while a controlled test on the high-fidelity solver revealed a quantifiable performance drop ($\sim 10^{-3}$ residual error) due to plant--model mismatch, underscoring the importance of surrogate accuracy and actuator modeling in practical deployment.
\color{black}

An important modeling challenge arises from the fact that rich closed loop data are often unavailable. In such cases, controllers must be designed based on the learned autonomous (possibly perturbed) dynamics alone, and the effect of actuation has to be inferred or approximated. While the underlying PDE evolves in continuous time, the NO surrogate provides only a discrete-time model, forcing us to reconcile controller design at the continuous-time level with its sampled-data implementation.

Looking forward, several challenges and exciting research directions emerge. Beyond the accuracy of the NO surrogate itself, a critical issue is the uncertainty in how actuators affect the dynamics. Future work should therefore incorporate uncertainty quantification not only for model error but also for actuator effects, for example through strategies such as wash-out filters~\cite{siettos2012equation, patsatzis2023data}. Extending the framework toward adaptive and model-predictive control, as well as output-feedback settings where the full state is not available, are natural and necessary directions. Ultimately, the real test of this approach will be its application to high-dimensional systems in the presence of only experimental data, where first-principles models are unavailable and both open-loop and actuator uncertainties must be actively managed. 









\addtolength{\textheight}{-2.1cm}

\end{document}